\documentclass[prl,reprint,doublecolumn,showpacs]{revtex4-1}
\usepackage{amsmath}
\usepackage{graphicx}

\begin{document}
\title{Boundary Curvature Effect on Thin-film Drainage and Slip Length Measurements}
\author{Angbo Fang}
\address{Department of Physics, \\Hong Kong University of Science and Technology,
\\Clear Water Bay, Kowloon, Hong Kong, China }
\date{\today}

\mbox{}

\begin{abstract}
 The thin-liquid film drainage between two curved surfaces is a fundamental process for many hydrodynamic measurements,
for which Vinogradova's formula has played a central role when flow slip occurs at fluid-solid interfaces.
By performing a rigorous order-of-magnitude analysis, we reveal the importance of the curvature contribution to boundary flow, neglected sofar. Vinogradova's result is found to considerably underestimate the slip-induced reduction of the hydrodynamic drainage force. Our theory can play a crucial role in distinguishing finite-slip from no-slip and quantifying the degree of flow slip at fluid-solid surfaces, which is a fundamental but controversial issue in fluid dynamics.  Moreover, qualitatively different from previous theories, our theory predicts a finite hydrodynamic repulsive force for two hydrophobic particles in touch, thus allowing particle collision to occur in a finite time without any additional attractive surface forces.  This finding has deep and immediate implications on  particle coagulation, adsorption and sedimentation processes relevant for numerous industrial technologies as well as natural phenomena on the earth.
\end{abstract}

\maketitle
The no-slip HBC~\cite{Lauga:2007}, assuming zero relative flow velocity adjacent to a solid surface, has served as an unquestionable pillar for fluid dynamics for more than one hundred years.  With the rapidly rising power of molecular dynamic simulations and increasing experimental capabilities for probing flow phenomena at micro- and nanometer scales,  it has undergone intensive challenges in recent years~\cite{Barrat:2007, Troian:1997, Vinogra:1999, Neto:2005}. It is now widely accepted that flow slip generally occurs near smooth hydrophobic surfaces.  The role of flow slip has become more and more prominent due to the recent blossoming of micro- and nanofluidic applications~\cite{Squires:2005, Whitesides:2006,Schoch:2008,Bocquet:2010} as well as the emergence of superhydrophobic surfaces~\cite{Lafuma:2003, Ma:2006, Roth:2010}.
The thin-film drainage problem~\cite{Chan:1985}, crucially involved in many processes including particle coagulation, wetting and
lubrication, film stability, as well as measurements of surface forces and slip lengths, has to take into account finite slip effects for hydrophobic surfaces.  The seminal work by Vinogradova~\cite{Vinogra:1995} provides an analytic formula for the slip-correction factor to the hydrodynamic force during the drainage of a thin liquid film between two hydrophobic spheres.
Her formula has been widely used in interpreting state-of-the-art
hydrodynamic force measurements using atomic force microscope (AFM)~\cite{Butt:2005,Vinogra:2003} or surface force apparatus (SFA)~\cite{Georges:1993, Claesson:1996} and especially, determining the slip lengths~\cite{Granick:2001, Granick:2002, Craig:2001, Bonaccurso:2002, Bonaccurso:2003, Charlaix:2003, Charlaix:2005, Rieutord:2004, Honig:2007, Maali:2008, Bhushan:2009, Mcbride:2009, Craig:2009, Ducker:2013} for various fluid-solid surfaces.  However, it suffers from the serious inadequacy of improperly discarding the boundary curvature effect, which is in fact quite important. 

In this letter we revisit the thin-film drainage problem and 
obtain new expressions for the hydrodynamic drainage forces, highlighting the boundary curvature effects.  We find Vinogradova's results  considerably underestimate the slip-induced reduction of hydrodynamic forces. As a consequence, many experiments based on drainage processes have to be reinterpreted. Remarkably, the boundary curvature effect also leads to a finite hydrodynamic repulsive force for two hydrophobic particles in touch, which is expected to have deep implications on particle coagulation processes.

Consider two undeformable spherical particles immersed in a Newtonian liquid with viscosity $\eta$ (Fig.~1).
The slip lengths at the sphere-fluid interfaces are given by $b_1$ and $b_2$. The distance between particles, $h$, is assumed to be small compared with both $R_1$ and $R_2$.
We define the weighted radius by $1/R_e=1/R_1+1/R_2$. 
\begin{figure}[ht!]
\centering
\includegraphics [width=2.1 in]{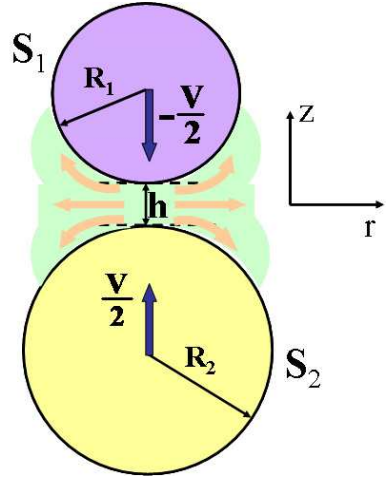}
\caption {(Color online) Drainage of a thin liquid film between two spheres. The sphere radii are $R_1$ and $R_2$, respectively. The distance of closest approach between the two surfaces is h and the velocities of the upper and lower spheres are respectively given by $-V/2$ and $V/2$.
The appropriate circular cylindrical coordinate system,  $(z, r)$, is also shown.}
\end{figure}
The Reynolds theory of lubrication is of zeroth order with respect to the small parameter $\epsilon=h/R_e$. The axisymmetric nature of the system geometry and fluid motion can be employed using a circular cylindrical system $(z, r)$ with the origin located at the midpoint of the gap.
The flow velocity components along the z axis and r axis are respectively given by $v_z$ and $v_r$.
The velocities of the upper and lower spheres are along the z axis and respectively given by $-V/2$ and $V/2$. 
Our interest will be focused on the inner region defined by $r<<R$, or, more precisely, the neighborhoods of the poles at $(z,r)=(\pm h/2, 0)$, which contributes the dominant part of hydrodynamic forces.   To better describe the inner region we introduce the ``stretched" inner variables~\cite{Brenner:1967} $\tilde{z}=z/h,  \quad \tilde{r}=r/\sqrt{hR_e}$, with which the upper and lower sphere surface are respectively described by
$\tilde{z} = \frac{1}{2} + (R_e/2R_1) {\tilde{r}}^2 +\frac{1}{8} \epsilon ({R_e}/{R_1})^4 {\tilde{r}}^4 + O(\epsilon^2)$ and $\tilde{z}= -\frac{1}{2} - (R_e/2R_2) {\tilde{r}}^2 +\frac{1}{8} \epsilon({R_e}/{R_2})^4 {\tilde{r}}^4 + O(\epsilon^2)$.   Introducing the rescaled dimensionless quantities,
$\tilde{v}_z=v_z/V$,  $\tilde{v}_r = \sqrt{\epsilon}v_r/V$ and  $\tilde{p}= \epsilon^2 p/(\eta V R_e)$, we obtain the reduced hydrodynamic equations: ${\partial^2 \tilde{v}_r}/{\partial \tilde{z}^2}={\partial \tilde{p}}/{\partial \tilde{r}}$  and ${\partial \tilde{p}}/{\partial \tilde{z}}=0$.  The continuity equation is given by
$\partial \tilde{v}_z/\partial \tilde{z} + \frac{1}{\tilde{r}}\partial (\tilde{r}\tilde{v}_r)/\partial \tilde{r}=0$.

To zeroth order of $\epsilon$, we have the hydrodynamic boundary conditions [see supplementary material]
\begin{equation}
\tilde{v}_z -  (R_e/R_1)\tilde{r}\tilde{v}_r = -\frac{1}{2};  \qquad -\frac{\partial \tilde{v}_r}{\partial \tilde{z}} -\frac{1}{\tilde{b}_1}\tilde{v}_r=0
\end{equation}  
at the upper sphere surface and
\begin{equation}
\tilde{v}_z + (R_e/R_2)\tilde{r} \tilde{v}_r= \frac{1}{2};  \qquad \frac{\partial \tilde{v}_r}{\partial \tilde{z}} -\frac{1}{\tilde{b}_2} \tilde{v}_r=0
\end{equation}    
at the lower sphere surface, with $\tilde{b}_1=b_1/h$ and $\tilde{b}_2=b_2/h$.
The curvature-induced renormalization of slip lengths~\cite{Liu:1990} is of order $\epsilon^1$ and discarded.  Remarkably, the curvature contribution to boundary flow normal to the sphere surfaces is of order $O(1)$ and can not be discarded.  

Integrating the continuity equation across the gap, we obtain for the pressure function $\tilde{p}$ [see supplementary material]:
\begin{equation}
-\frac{1}{2} \frac{1}{\tilde{r}} \frac{d}{d\tilde{r}}\left[ X \tilde{r} \frac{d\tilde{p }}{d\tilde{r}}\right]
=  6  + 3 b^* Z \frac{Z+ 2\overline{b}}{Z+\tilde{b}_1+\tilde{b}_2} \tilde{r}\frac{d\tilde{p}}{d\tilde{r}},
\end{equation}
where $Z=1+\tilde{r}^2/2$, $b^* = (\tilde{b}_1  R_2 +\tilde{b}_2 R_1)/(R_1+R_2)$, $\overline{b}=\tilde{b}_1\tilde{b}_2/b^*$,
and $X= Z^2(Z+B^+)(Z+B^-)/(Z+\tilde{b}_1+\tilde{b}_2)$ with $B^{\pm}= 2(\tilde{b}_1+\tilde{b}_2 \pm \sqrt{\tilde{b}_1^2 +\tilde{b}_2^2 -\tilde{b}_1\tilde{b}_2})$.  Notably, eq.~(3) differs from Vinogradova's pressure equation by the presence of the second term on the right hand side. This term arises from the curvature contribution to boundary flow normal to the fluid-sphere interfaces.  Importantly, it is of the order of $O(1)$ if $b^*$ is not negligibly small.  Therefore,  for many cases of interest, when either the slip length is moderately large or the gap width is small,  we should be keep this essential term.  

If $b^* \ne 0$, in general we cannot analytically solve the pressure equation. Here we focus on two important cases: ({\bf 1}) a hydrophobic sphere approaching a hydrophilic plate ($R_2\to \infty$ and $b_2=0$); ({\bf 2}) two spheres with similar surface treatments ($b_1=b_2\equiv b$).  Both cases are frequently encountered in hydrodynamic force measurements using AFM or SFA.
Case ({\bf 2}) describes the drainage between two similar spheres or crossed cylindrical surfaces
and is also particularly relevant for particle coagulation processes.  We should distinguish case ({\bf 1}) from case ({\bf 1'}) for a hydrophilic sphere ($b_1=0$) approaching a hydrophobic plate ($R_2\to \infty$ and $b_2 > 0$). Interestingly, for this case ($b^* \sim \epsilon $), the curvature effect is negligible for the lubrication approximation of the order of $\epsilon^0$ and our result would reduce to Vinogradova's, which does not distinguish case ({\bf 1}) from case ({\bf 1'}).

For case ({\bf 1}),  we have $b^*=\tilde{b}_1\equiv \tilde{b}$ and $\overline{b}=0$.
Eq.~(3) can be partially solved to give
\begin{equation}
\frac{d\tilde{p}}{dZ} = -6\frac{Z+\tilde{b}}{(Z+4\tilde{b})Z^3}
\left[1 - \beta  g_1(Z)\right]
\end{equation}
with $\beta =4\tilde{b}\equiv4b/h$,
$g_1(x)=3(1+\beta/x)^{3/2}\ln\left(\frac{\sqrt{x+\beta}+\sqrt{x}}{\sqrt{1+\beta}+1}\right)/(x-1)
-(1+\beta/x)(\sqrt{1+\beta}+2\beta/\sqrt{1+\beta})/(\sqrt{1+\beta}x+\sqrt{x^2+\beta x})
-1/x$. The hydrodynamic force acting on the sphere is [see supplementary material]:
\begin{equation}
F = (12\pi \eta R_e V) \frac{R_e}{h} \int^{\infty}_{0} d\tilde{r}^2 \int^{\infty}_{1+\tilde{r}^2/2}dZ \left(-\frac{d\tilde{p}}{dZ}
\right) = F_0 (f^*_v - f^*_c)
\end{equation}
where $F_0=6\pi \eta R_e^2 V/h$ is the force at the no-slip limit, $f^*_v = (1/4)(1+6/\beta[(1+1/\beta)\ln(1+\beta)-1])$ is Vinogradova's slip-correction factor, and $f^*_c = 2\beta \int^{\infty}_{0} dx \int^{\infty}_{x}dy g_1(1+y)$ is the curvature-correction factor.

\begin{figure}[ht!]
\centering
\includegraphics [width=2 in]{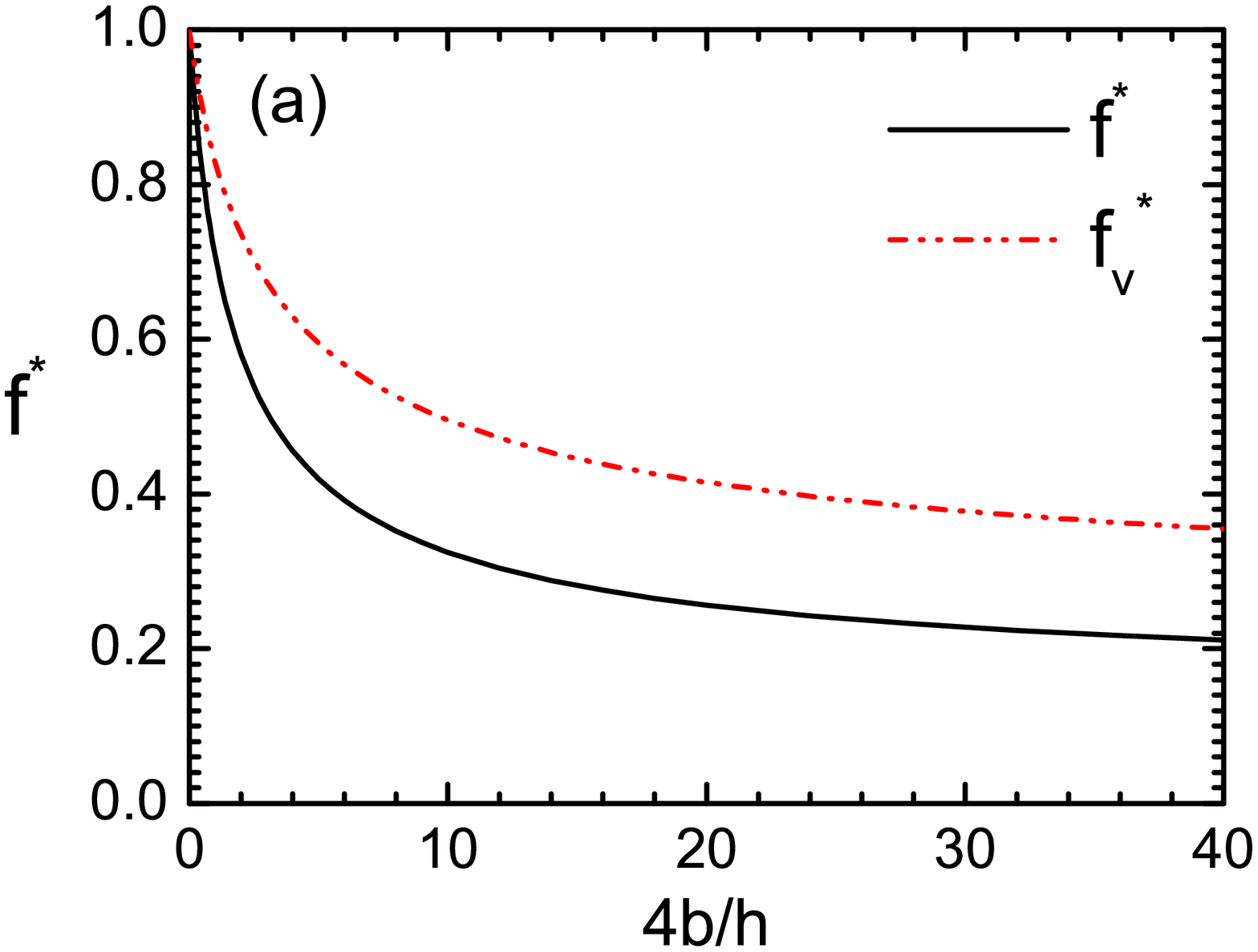}
\includegraphics [width=2 in]{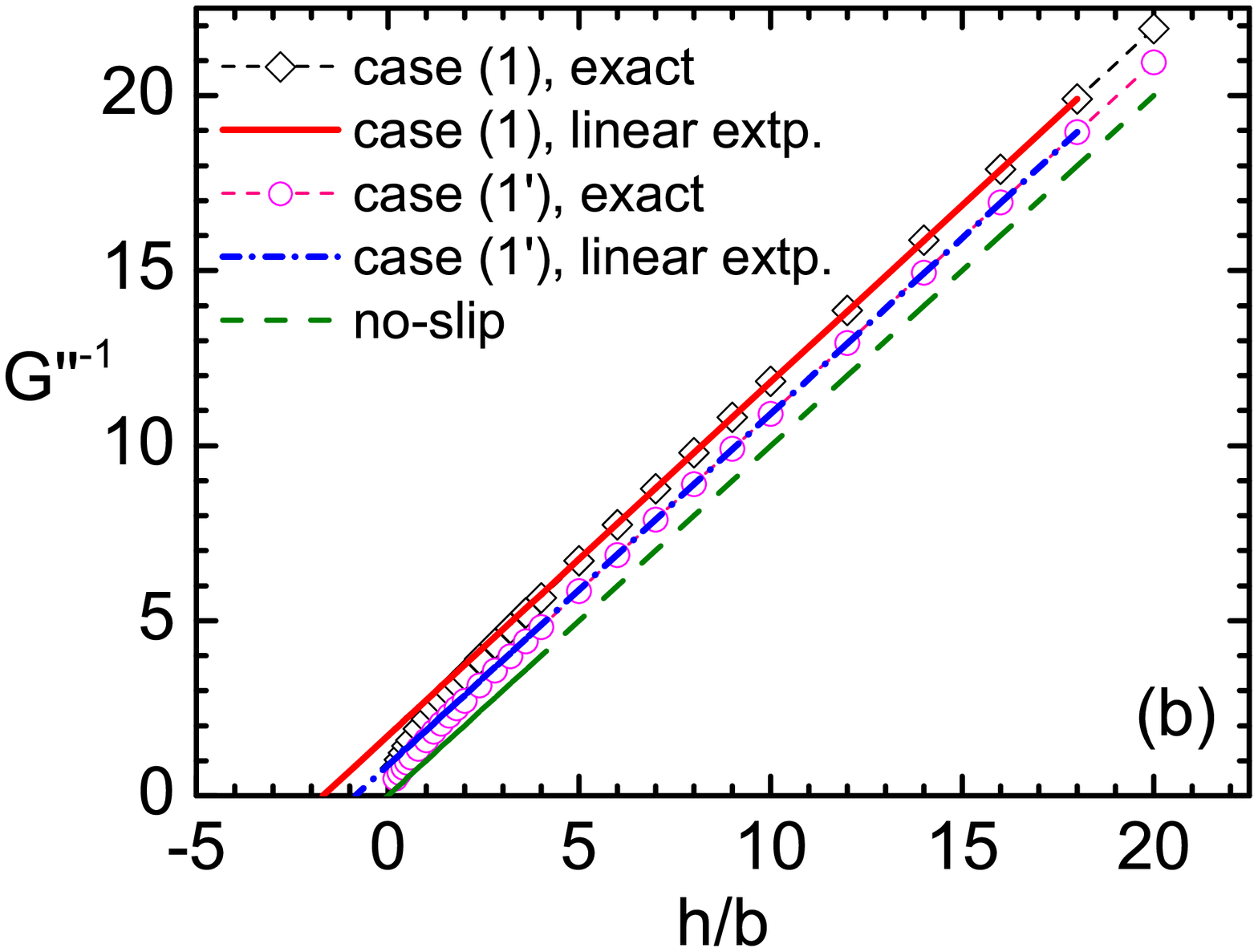}
\caption {(Color online) Slip-reduction of the hydrodynamic drainage force.
(a) Comparison for the correction factor of our theory with Vinogradova's with and without including
the boundary curvature effect, respectively.  (b) The inverse viscous damping $G''^{-1}$ (in units of
$1/6\pi\eta \omega R_e^2$) as a function of the gap width, for case (1) and (1').  The linear extrapolation lines are plotted to extract
the slip lengths by identifying the intercepts on the h axis. Due to the use of data in the range of $h/b \sim 8-18$, the intercepts for
case (1) and (1') are respectively given by $-1.7 b$ and $-0.9 b$, deviating from their respective asymptotic values, $-2b$ and $-b$.}
\end{figure}

When $b/h$ is small we may think the curvature effect is not important and Vinogradova's formula well captures the slip effect.
However, this is not true.  To first order of $b/h$ we have $f^*_v=1-b/h$ and $f^*\equiv f^*_v-f^*_c =1-2b/h$. Therefore, even if
$b\ll h$,  Vinogradova's formula underestimates the slip reduction of the hydrodynamic force by a factor of 2.  When $b$ is comparable to or large than $h$, the curvature correction should be even more important. Fig.~2a clearly shows that by neglecting of boundary curvature effect Vinoggradova's formula does not describe the slip effect sufficiently.  This insufficiency becomes very prominent as the sphere-fluid
interface becomes more slippery or the gap width becomes smaller.

There is an important class of experiments~\cite{Charlaix:2005, Charlaix:2008, Maali:2012} performed with a dynamic SFA in which the oscillating hydrodynamic force is measured when the sphere-plane gap is vibrated with amplitude $d_0$ and frequency $\omega/2\pi$. The viscous damping due to flow is given by $G''(\omega)=6\pi \eta \omega R^2 f^*/h$. Therefore, for case ({\bf 1}) the linear extrapolation of $G''(\omega)^{-1}$ at sufficiently large $h$ should intersects the $h$ axis at $-2b$.  In contrast, for case ({\bf 1'}), where our theory reproduces
Vinogradova's formula for the slip-correction factor, the intercept is $-b$.  In Fig.~2b we compare case ({\bf 1}) and ({\bf 1'}) for the inverse viscous damping as a function of the gap width. The extrapolation line crosses the h-axis at approximately $-1.7 b$ for case ({\bf 1}) and $-0.9 b$ for case ({\bf 1'}), both of which deviate from the respective exact prediction by nearly $10 \sim 15\%$ due to the use of data from the regime with $h$ not sufficiently large. This should be taken into account if more accurate values of slip lengths are to be extracted.
\begin{figure}[ht!]
\centering
\includegraphics [width=2 in]{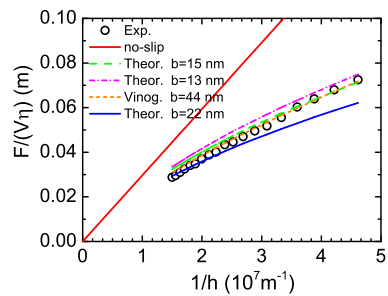}
\caption {(Color online) Comparison between theoretical predictions and experimental data for the hydrodynamic drainage force as a function of
the inverse gap width. The experimental data (extracted from fig.~3 of Ref.~\cite{Bonaccurso:2003}) is for the drainage of an aqueous sucrose solution (51.94 wt\%) between a hydrophobic sphere and a hydrophilic planar substrate.}
\end{figure}

To further demonstrate the importance of  the curvature effect, Fig.~3 is plotted to
compare theoretical predictions with the measured hydrodynamic drainage force as a function of $1/h$. Fitting with $f^*$ to the force curve shows that the experimental data in the whole probed range are well fitted by $b=15$ nm but at large $h$ the data agree better with our theoretical curve with $b=22$ nm.  It is hard to tell in which range the experimental data are more reliable, because noise-to-signal ratio is large for force measurement at large $h$~\cite{Attard:2011} while accurate separation measurement and control of approaching velocity are difficult at small $h$. In any way, this experimental situation belongs to case (1) and the full correction factor $f^*$ instead of $f^*_v$ should be used for theoretical fitting and extracting the desired slip length. By using Vinogradova's formula, Bonaccurso et. al. have obviously overestimated the slip length by a factor of $2\sim3$. Moreover, Fig.~4 is plotted to compare our theory with the measurement of hydrodynamic drainage force in di-n-octyl phthalate in a colloidal AFM experiment~\cite{Attard:2012}. Our theoretical prediction with $b=40$ nm agrees quite well with the experimental data except for the three smallest separations measured. On the other hand, by discarding the boundary curvature effect
Vinogradova's theory seriously underestimates the slip-induced reduction of hydrodynamic drainage force. Even with $b=100$ nm, it still considerably overestimates the drainage force for separations less than $70$ nm. The force at large separations ($>150$ nm) is insensitive to the magnitude of slip length, implying data in this range are not appropriate to determine the slip lengths. 
Therefore, we expect the separation range $h/b \sim 1-5$ plays a decisive role in accurately determining the slip length. In this regime, the boundary curvature effect is prominent
\begin{figure}[ht!]
\centering
\includegraphics [width=2 in]{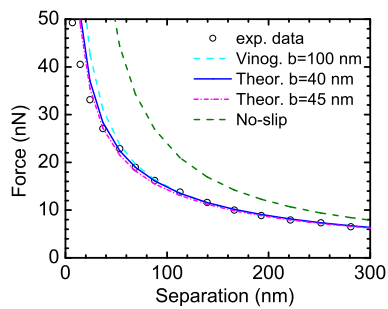}
\caption {(Color online) Comparison between theoretical predictions and experimental data for the hydrodynamic drainage force as a function of
the gap width. The experimental data (extracted from fig.~6(a) of Ref.~\cite{Attard:2012}) is for the drainage of di-n-octyl phthalate between an OTS silicon wafer and a silica microsphere. Fitting curves are shown for our theory with $b=40$ and $50$ nm and for Vinogradova's theory  with $b=100$ nm.}
\end{figure} 
and our full theory instead of Vinogradova's should be used for interpreting the experiments.

For case ({\bf 2}), we have identical slip lengths on the two spheres ($b^*=\tilde{b}_1=\tilde{b}_2\equiv\tilde{b}$).
The hydrodynamic force acting on the upper sphere is $F^s = F_0 f^*$, with
\begin{equation}
f^*= 1-2\beta_s\int^{\infty}_{0} dx \int^{\infty}_{x}dy  \frac{\ln[1+y/(1+\beta_s)]}{y(y+1)^3}
\end{equation}
where $\beta_s=6b/h$.
For $b\ll h$ we obtain $f^* = 1-3b/h+O((b/h)^2)$, substantially different from Vinogradova's corresponding result
$f^*_v =1-2b/h+O((b/h)^2)$.  Again, by discarding the curvature effect, Vinogradova has underestimated the slip reduction to hydrodynamic force
by a factor of 1.5, even at large gap widths.  On the other hand, with smaller gap width or more slippery surfaces the curvature effect becomes more important. Her formula becomes even less reliable and leads to considerable underestimation of the slip-reduction.  These observations call for reinterpretation of hydrodynamic drainage measurements~\cite{Granick:2001, Attard:2012} between two surfaces with identical slip lengths, with at least one of the surfaces is curved.

We plot in Fig.~5 the comparison between experimental data and theoretical fittings for a recent measurement of the hydrodynamic  drainage force in di-n-octyl phthalate~\cite{Attard:2012}. The experimental situation belongs to case ({\bf 2}). We first see that the measured force significantly deviates from the prediction by the Reynolds theory.  Second, at small to intermediate separation the experimental data can not be fitted well using Vinogradova's formula with $b=12$ nm proposed by the authors~\cite{Attard:2012}. Third, our theory can fit the experimental data very satisfactorily in the whole range of the measured separation, with the slip length predicted to be $16$ nm.  Fourth, for this specific driving velocity used in the experiment, there seems no need to introduce shear-rate-dependent slip length~\cite{Attard:2012}. Last but not least, with $b=24$ nm Vinogradova's formula can still fit the data rather well. This can be understood as follows.  The range of measured separation lies between $20\sim 500$ nm, for which the measured force is well described by $F_0 (1- D/h)$, with $D$ a constant to be determined.  Vinogradova's theory predicts $D=2b$, whereas our theory, taking into account the curvature effect, predicts $D=3b$. Thus both theories can fit the measured force in the whole measured separation range, but Vinogradova's formula leads to a slip length overestimated by $50\%$. Clearly, to observe the more prominent qualitative consequences of the curvature effect, we should push forward the experimentally probed regime to the range of separation smaller than the slip length.\begin{figure}[ht!]
\centering
\includegraphics [width=2 in]{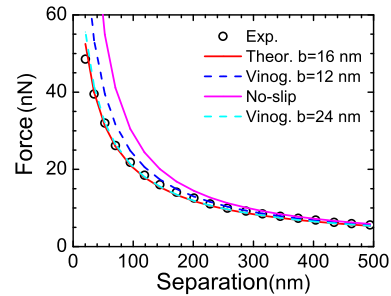}
\caption {(Color online) Hydrodynamic drainage force as a function of gap separation.
The experimental data is extracted from fig.~6(b) of Ref.~\cite{Attard:2012}.}
\end{figure} 
This puts stringent requirements on accurately measuring forces and separations and avoiding possible contamination when two surfaces are close to each other.  Within current experimental capabilities, it seems more easy to use spheres with slip lengths of the order of $100$ nm and carefully explore the separation range $20\sim 100$ nm.  For example, the boundary curvature effect should be of crucial importance to the hydrodynamic force for a bubble approaching a solid substrate~\cite{Manor:2008}.

We consider the other asymptotic limit $h/b \to 0$, which is particularly important for the coagulation process of hydrophobic particles.  Vinogradova's formula yields the hydrodynamic force as $F^s_v \sim (6\pi \eta V R_e^2) (1/3b) \ln(6b/h)$, which diverges as the particles get into touch, essentially not quite different from the lubrication force predicted by the Reynolds theory with no-slip HBC.  Nevertheless, by correctly including the curvature effect, our theory predicts (see supplementary information)
\begin{equation}
F^s \sim (6\pi \eta V R_e) \frac{R_e}{0.45 b} \qquad  \mbox{ as }  h\to 0,
\end{equation}
which remains finite even when the particles are in touch. Interestly, this means that even if the two particles are visually in touch ($h=0$), their mutual repulsive force can be described by the Reynolds theory with a finite effective separation given by $0.45 b$.
Remarkably and importantly, our theory, qualitatively different from either the Reynolds theory (diverging according to  $1/h$  as $h \to 0$)
or Vinogradova's result (diverging according to $-\ln{h}$ as  $h \to 0$),  allows the particle collision to occur in a finite time without any additional attractive surface forces.  This will have important implications for particle coagulation, adsorption and sedimentation relevant for many industrial and natural processes.

\section*{Acknowledgements}
I thank the hospitality of physics department of the Hong Kong University of Science and Technology where part of this work was done.
I acknowledge the support from King Abdullah University of Science and Technology (KAUST) via Award No. KAUST08/09.SC01.

\bibliographystyle{apspre}

\end{document}